\documentclass[sigconf]{acmart}

\usepackage{booktabs} 
\usepackage{todonotes}

\setcopyright{none}
\setcopyright{rightsretained}

\acmDOI{}
\def\acmDOI#1{\def\@acmDOI{#1}}
\acmISBN{}

\acmConference[LBRS@RecSys'18]{Late-Breaking Results track part of the Twelfth ACM Conference on Recommender Systems}{ October 2-7, 2018}{Vancouver, BC, Canada}
\acmYear{2018}
\copyrightyear{2018}

\begin{document}
\title{Who is Really Affected by Fraudulent Reviews?}
\subtitle{An analysis of shilling attacks on recommender systems in real-world scenarios}

\author{Anu Shrestha}
\affiliation{%
  \institution{Boise State University}
}
\email{anushrestha@u.boisestate.edu}

\author{Francesca Spezzano}
\affiliation{%
  \institution{Boise State University}
}
\email{francescaspezzano@boisestate.edu}

\author{Maria Soledad Pera}
\affiliation{%
  \institution{Boise State University}
}
\email{solepera@boisestate.edu}


\begin{abstract}
We present the results of an initial analysis conducted on a real-life setting to quantify the effect of shilling attacks on recommender systems. We focus on both \textit{algorithm performance} as well as the \textit{types of users} who are most affected by these attacks. 
\end{abstract}

%


\maketitle

\section{Introduction}
The effect of shilling attacks on recommender systems, where malicious users create fake profiles so that they can then manipulate algorithms by providing fake reviews or ratings, have been long studied. Previous work has characterized and modeled shilling attacks on recommenders, defined new metrics to quantify the impacts of these attacks on known recommender algorithms, and applied a \emph{detect+filtering} approach to mitigate the effects of spammers on recommendations (see recent survey~\cite{burke2015robust}). We observe from the literature that empirical analysis thus far has focused on assessing the robustness of recommender systems via \textit{simulated attacks}~\cite{burke2015robust,seminario2013accuracy}. Unfortunately, there is lack of evidence on what is the impact of fake reviews or fake ratings in a \textit{real-world }setting.

We present a preliminary analysis conducted to understand the influence of fraudulent reviews on the recommendation process. We do this through an initial study on known datasets with gold standards in different domains and a commonly-used recommendation algorithm. Our goal is to shed light on the effect of this attack and identify gaps to be addressed in the future by seamlessly connecting 
recommender and data mining research, as the latter has a rich body of work when it comes to spam detection and prevention. 

%
%
%
%

\section{Analysis Framework}


{\bf Datasets.} We use two real-world datasets (Table~\ref{fig:datasets}) that offer information about fraudulent reviews, 
which we treat as ground truth.

{\em Yelp!}~\cite{MukherjeeV0G13}: This dataset consists of Yelp reviews from two domains: hotels (\textbf{YH}) and restaurants (\textbf{YR}). Yelp filters fake/suspicious reviews and puts them in a spam list. A study found the Yelp filter to be highly accurate~\cite{Bloomberg} and researchers have used filtered spam reviews as ground truth for spammer detection (e.g., ~\cite{RayanaA15}). Spammers, in our case, are users who wrote at least one filtered review. 

{\em Amazon}~\cite{McAuleyL13}: Here we consider reviews from two domains: beauty (\textbf{AB}) and health (\textbf{AH}). In this case, we define ground truth following the framework in ~\cite{FayaziLCS15}, which is based on helpfulness votes. 
Thus, we treat as a spammer every user who wrote at least one review in which he rated a product as 4 or 5 and has helpfulness ratio $\leq 0.4$. 

\begin{table}[t]
\begin{tabular}{@{}lcccc@{}}
\toprule
\textbf{Dataset} & \multicolumn{1}{l}{\textbf{Users}} & \multicolumn{1}{l}{\textbf{Items}} & \multicolumn{1}{l}{\textbf{Ratings}} & \multicolumn{1}{l}{\textbf{Spammers}} \\ \midrule
\textit{Amazon-Beauty} & 167,725 & 29,004 & 252,056 & 3.26\% \\
\textit{Amazon-Health} & 311,636 & 39,539 & 428,781 & 4.12\% \\
\textit{Yelp!-Hotel} & 5,027 & 72 & 5,857 & 14.92\% \\
\textit{Yelp!-Restaurant} & 34,523 & 129 & 66,060 & 20.25\% \\ \bottomrule
\end{tabular}
\caption{Summary of datasets}\vspace{-8mm}
\label{fig:datasets}
\end{table}

{\bf Experimental setting.} In this paper, we analyze the robustness to shilling attacks of matrix factorization (MF), a commonly-used recommender algorithm. 
We used probabilistic MF~\cite{mnih2008probabilistic} with 40 latent factors and 150 iterations. We performed 5-fold cross-validation and measured the performance in terms of \textit{RMSE}\footnote{Using \textit{hitRatio}, we obtained similar outcomes. Thus, due to space limitations, we excluded that metric from our discussion.} only for non-spam users. We also used \textit{prediction shift} (PS), a measure explicitly defined to quantify the impact of spammer attacks on recommenders, which captures the average changes in predicted ratings~\cite{burke2015robust}~\footnote{In addition to PS, we considered \textit{stability of prediction}, another common measure to quantify spammer attacks. As it is inversely proportional to PS, we only report PS.}.

\section{Results \& Discussions}
We discuss the effect of fake reviews on recommendations offered to users in \textit{real-world }scenarios, as opposed to \textit{simulated}  attacks.

{\bf Do spam ratings affect recommendations?}
By following the classical evaluation framework for shilling attacks on recommender systems~\cite{burke2015robust}, we measured the performances on the original dataset (with spam) and when we remove all the reviews written by spammers (shilling attack).
We report the results of our assessment in Table~\ref{fig:results}. 

We anticipated a lower RMSE when removing spam. However, we did not observe this trend among most datasets in our study. This result aligns with previous work reporting (simulated) shilling attacks are not detectable using traditional measures of algorithm performance~\cite{LamR04}. 
Previous works also show PS values ranging from 0.5 to 1.5 when shilling attacks are simulated. However, we observe very low values in real-world scenarios: in our case, considered, PS ranges from 0.047 to 0.15, which we argue is not enough to promote or demote products attacked by the spammers. 
%
We believe this to be one of the reasons why algorithm robustness is not reflected by average metrics like RMSE. 
Further, looking at users as a whole does not help us quantify how much spammers are able to deceit recommenders or who are the users that are affected the most.

\begin{table}[h]
\begin{tabular}{@{}lcc@{}}
\toprule
\textbf{Dataset} & \multicolumn{1}{l}{\textbf{\begin{tabular}[c]{@{}c@{}}W/ Spammers\\(RMSE, PS)\end{tabular}}} & \multicolumn{1}{l}{\textbf{\begin{tabular}[c]{@{}c@{}} W/o Spammers \\RMSE\end{tabular}}} \\ \midrule
\textit{Amazon-Beauty} & (0.871, 0.122) & 0.901 \\
\textit{Amazon-Health} & (1.056, 0.047) & 1.053\\
\textit{Yelp!-Hotel} & (1.124, 0.150) & 1.125\\
\textit{Yelp!-Restaurant} & (1.039, 0.133) & 1.034 \\ \bottomrule
\end{tabular}
\caption{RSME and PS on datasets with and without spam.}\vspace{-6mm}
\label{fig:results}
\vspace{-1.5mm}
\end{table}

\begin{figure}[t]
\includegraphics[width=0.45\textwidth]{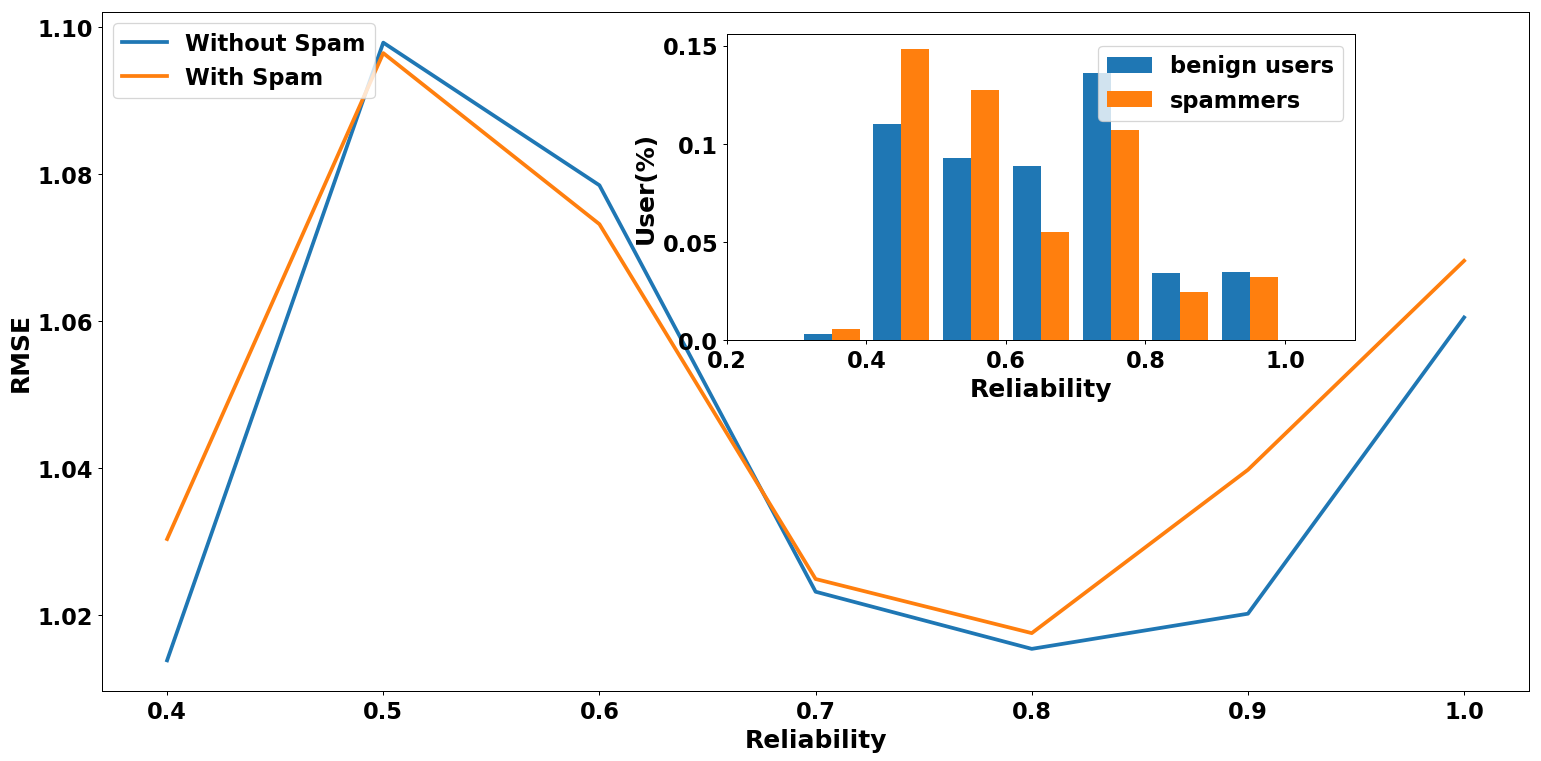}
\vspace{-3mm}
\caption{Yelp-Restaurants (YR); RMSE differences for ranges [0.4,0.5] and [0.7,1] are significant, $p<0.001$.}\label{fig:plotYelp}
\vspace{-4.25mm}
\end{figure} 
{\bf Who is really affected by spammers?}
%
To better understand 
which users are really affected by spammers, we analyzed users based on their \textit{reliability}: the ability of a user to rate a product according to what it deserves; as in  
~\cite{icdm16}. The rating a product deserves often aligns with what the majority of benign users think about that product given that they outnumber spam users.
Then, a \textit{reliable} user is one that always rates according to what a product deserves, whereas an \textit{unreliable} user deviates from that value. It seems, however, that by this definition spammers are unreliable as they may try to promote or demote a product by rating it differently than benign users. However, benign users may also be mistakenly treated as unreliable users, if they happen to be the type of users that disagree with average ratings assigned by the majority. 

\begin{figure}[t]
\includegraphics[width=0.45\textwidth]{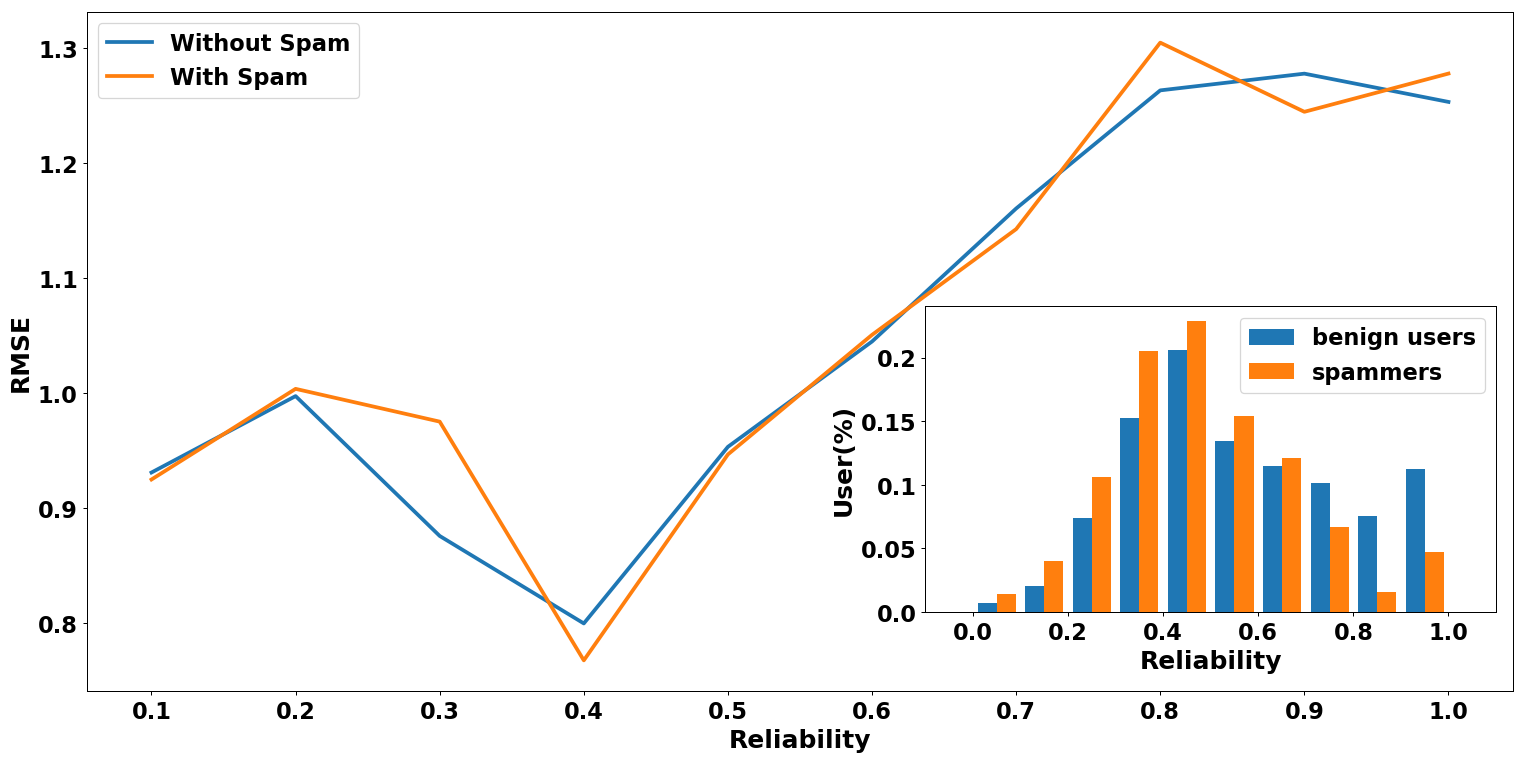}
\vspace{-3mm}
\caption{Amazon-Health (AH); RMSE differences for ranges [0.2,0.4] and [0.7-1] are significant $p<0.01$.}\label{fig:plotAmazon}
\vspace{-5.25mm}
\end{figure}

Figures~\ref{fig:plotYelp} and~\ref{fig:plotAmazon} illustrate how the RMSE varies according to the reliability of the users and the distribution of user reliability in YR 
and AH
\footnote{Results are similar for the other two datasets, excluded for space limitations.}. We observe that spammers and benign users have similar reliability distributions. 
Unfortunately, this is why spammers are able to camouflage as benign users very well, in terms of reliability. Regarding RMSE, we see that benign users that benefit from removing spam when generating recommendations are either \textit{unreliable} users or \textit{very reliable} ones. This result depends on spammers' reliability. Intuitively, the users more affected by spammers are the ones exhibiting a rating behavior very different from the spammers. Thus, \textit{traditional} spammers, i.e., unreliable ones, impact reliable benign users (right tail of the plots), while \emph{smart} spammers, i.e., the ones that are able to camouflage themselves as reliable benign users, affect unreliable benign users, i.e., those who disagree with the average (left tail). This latter result aligns with what observed for trust-based recommenders~\cite{golbeck2006generating}, which is not unexpected if we think 
of spammers as untrusted users in the network, in our case.

Overall, 
26.6\% (4\%, 5.6\%, 0.7\%) of benign users receive worse recommendations in presence of spam on YR (YH, AH, AB, resp.)
. We infer that the high (low, resp.) percentage observed for YR (AB, resp.) 
is due to the  proportion of spam in the dataset (see Table \ref{fig:datasets}). In a real-world scenario, the aforementioned percentages would translate into hundreds of thousands of users who would not be equally satisfied by recommenders that are not robust to shilling attacks. 
We believe this to be why this area warrants further study to make recommender algorithms not only \textit{more robust}, but also able to better serve \textit{all type of users} through stricter spam detection. 



\section{Conclusions \& Future Work}
We have presented the results of an initial empirical analysis that has allowed us to demonstrate that trends observed as a result of simulated shilling attacks on recommender algorithms remain the same in a real-world scenario. We validated that average metrics are not able to properly capture attack effect and that in the presence of spam, recommender algorithms are not uniformly robust for all type of benign users. 
These initial discoveries lead us to argue in favor of 
new algorithms that are not only robust to attacks, but that also ensure that all users are protected against spam while supporting spam detection that accurately spots the subset of spammers who in fact affect recommendations without mistreating non-traditional users (i.e., users whose taste differs from the popular) 
as spammers.

\bibliographystyle{ACM-Reference-Format}
\bibliography{paper-references}

\end{document}